\documentclass{aa}  
\usepackage{graphicx,times}
\newcount\lecurrentfam
\def\LaTeX{\lecurrentfam=\the\fam \leavevmode L\raise.42ex
\hbox{$\fam\lecurrentfam\scriptstyle\kern-.3em A$}\kern-.15em\TeX}

\begin{document}

\thesaurus{ 03
              (
               13.25.3;                
               09.03.2;                
               02.18.5                 
             ) 
              }

%
  \title{Nonthermal Hard X-Ray Emission from the Galactic Ridge} %

\authorrunning{V.A.\ Dogiel et al.}
\titlerunning{Nonthermal Hard X-Ray Emission from the Galactic Ridge}
\author {V.A.\ Dogiel\inst{1,2}, V.\ Sch\"onfelder\inst{1} , A.W.\ 
Strong\inst{1} }  
\institute{ Max-Planck Institut f\"ur 
extraterrestrische Physik, Postfach 1603, D-85740 Garching, FRG\\
   email: vos@mpe.mpg.de,  aws@mpe.mpg.de 
\and 
P.N.Lebedev Physical Institute, 117924 Moscow, Russia\\
email: vad@lpi.ru}  
\offprints{V.A.Dogiel \\ {E-mail: vad@lpi.ru}}  
\date{Received ?; accepted ?} 
  
   \maketitle  
  
\def\carbon{$^{12}$C\ } 
\def\oxygen{$^{16}$O\ } 
\def\pio{$\pi^o$-decay\ } 
 
\begin{abstract}  
We investigate the origin of the nonthermal X-ray emission  from 
the Galactic ridge in the range $10-200$ keV.  We consider 
bremsstrahlung of subrelativistic cosmic ray protons and 
electrons as production processes. From the solution of the 
kinetic equations describing the processes of particle {\it in 
situ} acceleration and spatial propagation we derive parameters 
of the spectra for protons and electrons. It is shown that the 
spectra must be very hard and have a cut-off at an energy $\sim 
150-500$ MeV for protons or $\leq 300$ keV for electrons. For 
{\it in situ } acceleration the flux of accelerated particles 
consists mainly of protons since the ratio of the accelerated 
protons to electrons  is large and the flux of nuclei with 
charges $Z>1$ is strongly suppressed. We show that the gamma-ray 
line flux generated by protons does not exceed the upper limit 
derived from observations if we assume that the X-ray ridge 
emission is due to  proton bremsstrahlung. However, the flux of 
$\pi^o$ photons produced by the accelerated protons is higher 
than the observed flux from the Galactic ridge if the cut-off is 
exponential for $\geq 150$ MeV.  If  the cut-off in the 
spectrum is extremely steep its value can be as large as $400$ 
MeV, just near the threshold energy for $\pi^o$ photon 
production. In this case the flux of gamma-rays is negligible 
but these protons  still  produce X-rays up to  $200$ keV. If 
a significant part of the hard X-ray emission at energies $\sim 
100$ keV is emitted by unresolved sources, then the energy  of 
X-rays produced by the protons does not have to exceed several 
tens keV.  Therefore, the cut-off energy can be as small as 
$30-50$ MeV and in this case the flux of $\pi^o$ photons is 
negligible too.  But for small cutoff energies  the flux of 
nuclear gamma-ray lines exceeds significantly the upper limit 
derived from the COMPTEL and OSSE data. Hence the cut-off of the 
proton spectrum has to be somewhere in between $50-150$ MeV in 
order not to exceed both $\pi^o$ and gamma-ray line fluxes.  
However the energy density of the CR protons would have to be 
$\sim 400$ eV cm$^{-3}$ which seems implausible. 
 
If on the contrary the hard X-ray emission from the disk is 
emitted by  accelerated electrons we do not have the problems of 
gamma-ray line and $\pi^o$ fluxes at all, and the required 
energy density of particles is only $\sim 0.2$ eV cm$^{-3}$. But 
in this case we must assume that acceleration of protons is 
suppressed. We discuss briefly the  possible origin of this effect. 
 
We have also estimated the ionization rate produced by the 
accelerated particles in the interstellar medium, and it is 
found that ionization of the medium would  be very significant 
for both energetic protons and electrons. In this way we may 
perhaps  resolve the problem of the observed large ionization 
rate.

 
\end{abstract}

\thesaurus{  
               03  
              (13.07.02;               
               11.01.2;                
               11.17.4);  
              }  
%
  
%

\section{Introduction}  
The diffuse emission from the Galactic ridge has been observed 
over a very broad energy range - from the radio to the gamma-ray 
band.The origin of the radio and gamma-ray emission is more or 
less clear \footnote{ Though the excesses of the diffuse 
gamma-ray emission  observed in the energy ranges below 30 MeV 
(Strong et al. 2000) and above 1 GeV (Hunter et al. 1997) are 
inconsistent with the standard parameters used in models of 
cosmic ray origin}. It is mostly nonthermal and produced by 
relativistic electrons and nuclei which interact with the 
interstellar magnetic field, background gas and low energy 
photons.    An important point is that the cosmic ray (CR) 
luminosity needed to produce this emission can be derived 
directly from observational data, and it is almost model 
independent (see Berezinskii et al. 1990) 
\begin{eqnarray} 
\label{lcr} 
L_{cr}&&\sim {W_{cr}\over t_{cr}}={{c w_{cr}\rho V}\over{\rho c t_{cr}}}=\\ 
&&{{c w_{cr}M_g}\over x}\sim 5\cdot 10^{40}~\mbox{erg s$^{-1}$}\,. 
\nonumber 
\end{eqnarray} 
Here: 
 
$w_{cr}$ is the CR energy density measured near Earth; 
 
$V$ is the volume of the Galaxy filled with CRs;

$W_{cr}\sim w_{cr}V$ is the total  energy of     Galactic CRs: 
 
$t_{cr}$ is the CR lifetime; 
 
$\rho$ is the average gas density in the Galaxy; 
 
$M_g\sim \rho V$ is the total gas mass in the Galaxy  determined from 
radio data; 
 
$x\sim \rho c t_{cr}$ is the grammage traversed by cosmic rays in the Galaxy 
which is determined from the CR chemical composition near Earth. 
 
The necessary energy for CR production can be supplied by 
supernovae (SN)  (see Ginzburg and Syrovatskii 1964). Indeed, 
the average energy release of a supernova explosion is about of 
$10^{50-51}$ erg,  which occurs every $30-100$ years in the 
Galaxy. Then the average supernova power is about 
\begin{equation} 
L_{SN}\sim 10^{41-42}~\mbox{erg s$^{-1}$} 
\label{lsn} 
\end{equation} 
Therefore $\sim 10$\% of the explosion energy should be transformed 
into the energy of fast particles to maintain the CR Galactic 
luminosity. This can be realized in processes of particle 
acceleration by SN shocks (see, e.g., Berezhko et al. 1994). 
This model of CR production by SNs successfully describes the 
characteristics of the diffuse radio and gamma-ray emission from 
the Galaxy (Berezinskii et al. 1990, see also Strong et al. 
2000). Indeed, the total energy flux of gamma-ray emission produced by 
the proton CR component in the galactic disk is $L_\gamma\sim 10^{39}$ erg sec$^{-1}$. 
From this value the necessary flux of CRs $L_{cr}$ can be estimated from 
 
\begin{equation} 
L_{cr}\sim L_\gamma {{x_\gamma}\over x}\,, 
\label{xg} 
\end{equation} 
 
where $x_\gamma$ is the matter thickness needed to produce a gamma-ray photon, 
$x_\gamma\simeq 120$ gr cm$^{-2}$, and $x$ is the grammage traversed by CRs 
in the Galaxy, $x\simeq 12$ gr cm$^{-2}$. Then from Eq.(\ref{xg}) we obtain 
that the CR energy flux which produce galactic gamma-rays is $L_{cr}\sim 
10^{40}$ erg sec$^{-1}$ that is in agreement with the value (\ref{lcr}). 
 
However, attempts to extend this model to the region of the 
nonthermal hard X-ray band (below $100$ keV) have serious 
problems. According to measurements of the nonthermal spectrum 
of the diffuse emission by OSSE (Kinzer et al. 1999) and GINGA 
(Yamasaki et al. 1997) the nonthermal X-ray flux in the energy range 
above $10$ keV is $L_x \sim 10^{38}$ erg sec$^{-1}$. 
If  the  $10$ keV nonthermal 
emission is due to electron bremsstrahlung 
(see Skibo et al. 1997) the necessary energy flux can be estimated as 
\begin{equation} 
L_{cr}\sim L_x {{\tau_x}\over \tau_i}\,, 
\label{tg} 
\end{equation} 
where $\tau_x$ is the characteristic time for production of 
bremsstrahlung photons and $\tau_i$ is the characteristic life-time of the 
electrons due to ionization losses. 
The ratio $\tau_x /\tau_i \sim 10^5$ then from Eq.(\ref{tg}) we obtain that 
the CR luminosity  is about $10^{43}$ erg s$^{-1}$ 
This is much more than the value (\ref{lcr}) and even more than that 
 can be supplied by SNs (\ref{lsn}). 
 
In this respect it is reasonable to 
 assume that there may be  two sources of CRs in the Galaxy: SNR 
 which produce particles in the relativistic   energy range  up 
to $10^{14}-10^{17}$ eV and another unknown very powerful source 
which generates subrelativistic particles. Below we discuss this 
problem in more detail. We start from the observational data.  
 
\section{Observational data} 
\begin{figure}[thbp]  
\mbox{
\includegraphics[width=0.7\hsize,angle=90,clip]{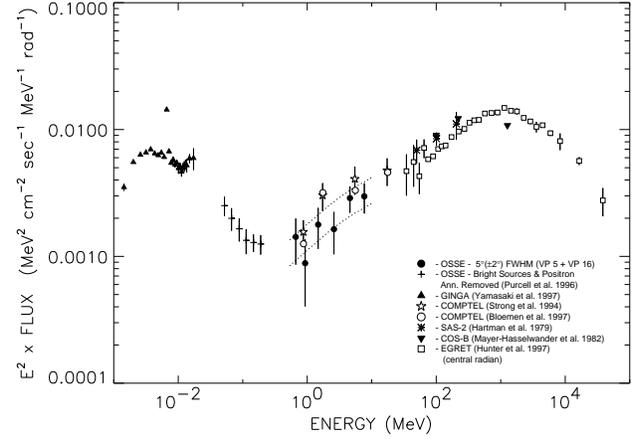}
}
 \caption[]{The  spectrum of diffuse emission from the Galactic disk (this figure was taken from Kinzer et al. 1999).  

\label{osse}} 
\end{figure}

Analysis of the observed X-ray flux in the range $2 - 16$ keV 
with the GINGA satellite (Yamasaki et al. 1997) showed that 
there is a hard component in the ridge spectrum in addition to 
the hot plasma component (see Fig.\ref{osse}). The total estimated luminosity is 
around $2\cdot 10^{38}$ erg s$^{-1}$ in the $3 - 16$ keV energy 
range. The combination of the GINGA spectrum with  measurements 
at higher energies  shows that the emission spectrum can be 
represented as a power-law over a very broad energy band 
without any flattening in the low energy range due to 
ionization losses. This means that the X-ray flux is produced 
in regions where the electrons are still freshly accelerated.  
 
Observations with the RXTE telescope also show a  hard X-ray 
excess above the thermal emission (Valinia and Marshall 1998). 
 
Analysis of the ASCA data (the energy range $0.5-10$ keV) led to the 
 conclusion that this emission cannot be due to unresolved 
point-like sources since  a class of sources with the required 
properties is not known and in fact can be excluded (Tanaka et 
al. 1999). Hence  the emission is most likely of diffuse 
origin.  
 
The diffuse flux can be produced either by emission by a hot 
plasma with a temperature $>7$ keV or  by fast (nonthermal) 
particles. The thermal origin of the emission seems  to be 
doubtful since: 1) it is not clear how to explain the plasma 
confinement in the disk because the thermal velocity in this 
case exceeds significantly the  escape velocity from the 
Galactic plane,  2) if the plasma is heated by supernova 
explosions, this assumption requires  a too large SN explosion 
rate of one per several years (see, e.g., Yamasaki et al. 
1997), 3) the width of X-ray lines observed in the direction of 
the Galactic ridge far exceeds that expected  for the case of 
thermal broadening (Tanaka et al. 2000). All these facts cast 
doubt on a thermal origin of the observed X-ray spectrum. 

 Another candidate for hard X-ray production in the Galactic ridge could be extended discrete sources  like  supernova remnants (SNRs). If the angular size of these remnants exceeds $1^o$, an ensemble of these SNRs may overlap with each other and form a smooth brightness distribution indistinguishable from  truly diffuse emission. Analysis of the SNR bremsstrahlung X-ray emission has been performed by Baring et al. (2000) who showed that a significant flux of hard X-rays is generated  by electron bremsstrahlung in a so-called ``Coulomb halo'' around  SNR whose maximal extent for 24 keV electrons is 100 pc.  
This shell-connected 
emission dominates the volume-integrated emission from SNR environs 
out to 100pc from the centre of SNRs. If the density of SNRs were high enough (the filling factor should be higher than $10^{-4}-10^{-3}$) the truly diffuse interstellar emission would be obscured by the emission of the SNR X-ray halos. In this case the major contribution to unresolved diffuse X-ray emission of the Galactic ridge would come from these discrete extended sources. However 
, the most optimistic estimates of the SNR filling factor lead to the lower part of this range. Thus, Koyama et al. (1986) showed that the SNR remnant scenario of the hard X-ray origin requires an unbelievably high supernova frequency, as high as  one every 10 years (see also Yamauchi et al. (1996) and Kaneda et al. (1997)). The flux of an individual ``Coulomb halo'' for the total ridge flux $2\cdot 10^{38}$ erg s$^{-1}$ in the energy range $2-16$ keV is of the order $\geq 10^{35}$ erg s$^{-1}$ for about one thousand SNRs in the central region of the galactic disk.  
However, there have been only a few detections of SNRs having such a very high temperature ($>2$ keV) and a such high luminosity ($\sim 10^{35}$ erg s$^{-1}$. The recent CHANDRA results have confirmed that the hard X-ray emission of the Galactic ridge is truly diffuse (Ebisawa et al. 2001). 
 
Hence, among several candidates for the hard X-ray Galactic ridge flux only the truly diffuse emission has remained.

Thus, from the observations of the GINGA, RXTE and ASCA 
telescopes it follows that the hard X-ray emission is diffuse 
and nonthermal at least up to  $16$ keV. 
 
Little  is also known about the origin of the diffuse ridge 
emission at higher X-ray energies. This energy range was 
investigated with  OSSE (Kinzer et al. 1999), WELCOME-1 
(Yamasaki et al. 1997) and RXTE (Valinia and Marshall 1998). A 
difficult task for these experiments is to distinguish the 
diffuse emission from that of unresolved sources. From the OSSE 
observations it follows that there are at least three 
components in the central ridge continuum spectrum below $1$ MeV: 
a variable soft component   which contains strong contributions 
from discrete sources; a positronium continuum; and a 
significant part of the emission which may be due to CR 
interaction with the gas and photons although there are 
certainly further  contributions from weaker  sources .  
The spectrum of the diffuse emission is flat  below $35$ keV. 
However above this energy there is a significant steepening in the 
spectrum. The  spectrum in the range $10-400$ keV is at best described 
by an exponentially cutoff power-law  of the form  (Valinia et 
al. 2000a)  
\begin{equation} 
I_x\propto E_x^{-0.6}\exp(-E_x/40~\mbox{keV})\,, 
\label{sp_val} 
\end{equation} 
i.e. the spectrum of the hard X-rays cuts off  above $40$ keV. 
Thus, it is reasonable to assume that there are at least three 
components of the diffuse emission in the Galactic disk: the 
hot background plasma  emitting the ridge thermal emission and 
peaked at $3$ keV, a process emitting hard X-rays  above $10$ 
keV, and SNRs which accelerate the Galactic cosmic rays 
generating the diffuse gamma-ray emission.  
 
 The existence of these  components  can be seen in 
 Fig.\ref{osse} which shows the total ridge spectrum for X-rays 
and gamma-rays. 
 
 *
 
The ridge X-ray emission in the range  $10-16$ keV is 
definitely diffuse and nonthermal. The origin of the emission 
at higher energies up to several hundred keV is not clear but 
there are  reasonable arguments that it is also diffuse and 
nonthermal although the fraction of the emission contributed by 
discrete sources is unknown. 
 
It is worth  mentioning that this excess of hard X-ray emission 
observed in the Galaxy is not unique. An excess of hard X-rays 
above the thermal flux has been observed in the spectra of the 
galaxy clusters  Virgo (Lea et al. 1981) and Coma 
(Fusco-Femiano et al. 1999); this has been interpreted as a 
result of particle {\it in situ} acceleration in the 
intracluster medium (see, e.g., En\ss lin et al. 1999, Dogiel 
2000, Sarazin and Kempner 2000).    
 
Various models for the nonthermal origin of the ridge excess 
have been presented. They include:  inverse Compton scattering 
 of relativistic electrons, bremsstrahlung radiation of 
 subrelativistic electrons or protons.  As Skibo et al. (1996) 
showed, however, the IC scattering of relativistic electrons 
cannot produce the bulk of hard X-ray emission below $100$ keV 
since these electrons generate also a   flux of radio emission 
which is higher than that observed.  In this paper we discuss 
two other mechanisms: electron bremsstrahlung and proton 
bremsstrahlung (or inverse bremsstrahlung). 
 
\section {The Energy and Particle Outputs in Emitting Particles}  
The bremsstrahlung photon production $Q_x$ by parent particles 
with  production rate $Q$ can be estimated as 
\begin{equation} 
Q_x\sim Q\cdot {\tau_{l}\over \tau_{br}}~\mbox{(particles sec$^{-1}$)}\,, 
\label{qx} 
\end{equation} 
where $\tau_l$ is the lifetime of the emitting particles which 
in our case is determined by Coulomb collisions (if the 
particles do not escape from the emitting region), and 
$\tau_{br}$ is the characteristic time for the production of 
X-ray photons by bremsstrahlung.  
 
The bremsstrahlung cross-section for production of a photon 
with  energy $E_x$ by electrons or protons can be written in 
the form (see e.g. Hayakawa 1969) 
\begin{eqnarray} 
  {{d\sigma_{br}\over{dE_x}}}&&={8\over 3}{Z^2}{{e^2}\over{\hbar c}}\left({{e^2} 
  \over{m{c^2}}}\right)^2{{m{c^2}}\over{E^\prime}}{1\over{E_x}}\cdot\\ 
&&\ln {{\left(\sqrt{E^\prime}+\sqrt{{E^\prime}-{E_x}}\right)^2}\over{E_x}}\,.  
\label{sbr} 
\nonumber 
\end{eqnarray} 
Here we should take $E^\prime=E_e$ for  electrons where $E_e$ 
is the electron kinetic energy,  and $E^\prime = (m/M)E_p$ for 
protons where $E_p$ is the proton kinetic energy,  since a 
photon with  energy $E_x$ can be produced either by an electron 
with  energy $E_e\sim E_x$ or by a proton with  energy $E_p\sim 
(M/m)E_x$; here $m$ and $M$ are the electron and proton rest 
masses (see in this respect estimations of expected X-ray fluxes from Orion produced by subrelativistic electrons and protons presented in Dogiel et al. 1997, 1998). 
 
The flux of bremsstrahlung photons is in this case  
\begin{equation} 
Q_x= \int\limits N(E)nv {{d\sigma_{br}(E,E_x)}\over{dE_x}}dE\,. 
\label{brf} 
\end{equation} 
The time $\tau_{br}$ from Eq.(\ref{qx}) is 
\begin{equation} 
\tau_{br}\sim n v \sigma_{br}\,, 
\label{br} 
\end{equation} 
where $n$ is the gas density and $v$ is the particle velocity. 
The value of $\tau_{br}$ is of the same order for the electrons 
and protons. The point is that the velocities of the electron 
with the energy $\sim E_x$ and of the proton with the energy 
$\sim (M/m)E_x$ are equal. For the same reason the rates of 
ionization loss  are also of the same order for these 
particles 
\begin{equation} 
  {\left({dE\over{dt}}\right)_i}=-{{2\pi {n}{e^4}}\over  
  {m{v}}}{\ln\Lambda_n}\,, 
\label{il} 
\end{equation} 
where $\ln\Lambda_n$ is a logarithmic function weakly dependent on the particle  
energy. But their lifetimes due to ionization losses  
\begin{equation}  
{\tau_i}\sim {{E_{e,p}}\over{(dE/dt)_i}} 
\label{taui} 
\end{equation} 
differ from each other by a factor  $(m/M)$. It follows from 
Eq.(\ref{qx}) that  protons with energy $(M/m)E_x$ produce many 
more photons with energy $E_x$ than  electrons with  energy 
$E_x$. Then for the same flux $Q_x$ we have  
\begin{equation} 
{Q_p\over Q_e}\sim {m\over M}\,. 
\end{equation} 
However, if we compare their energy outputs   
\begin{equation} 
F\sim E \cdot Q~\mbox{(erg sec$^{-1}$)} 
\end{equation} 
we find 
\begin{equation} 
{F_p\over F_e}= {{Q_pE_p}\over{Q_eE_e}}={{(m/M)Q_e (M/m)E_x} \over {Q_e E_x}} = 1\,. 
\end{equation} 
This means that the same energy flux in protons or electrons is 
necessary to produce a flux of X-ray photons $Q_x$. 
 
As follows from Eq.(\ref{brf}) the proton and electron 
bremsstrahlung fluxes are equal   if  the densities of emitting 
protons and electrons are equal 
\begin{equation} 
N_p({M\over m}E_x)\sim N_e(E_x)\,, 
\label{equal} 
\end{equation} 
 since the cross-sections and velocities are  the same for the 
electrons and protons. 
 
For a power-law differential spectrum of  particles $N=K^\prime 
E^{-\gamma^\prime}$ the   bremsstrahlung flux produced by the 
protons equals  that produced by the electrons if   
\begin{eqnarray} 
&&K_p^\prime {\int\limits_{m/ME_x}^{E_{max}^p}} E_p^{-\gamma^\prime}nv {{d\sigma_{br}(E_p,E_x)}\over{dE_x}}dE_p=\\ 
&&K_e^\prime {\int\limits_{E_x}^{E_{max}^e}} E_e^{-\gamma^\prime}nv {{d\sigma_{br}(E_e,E_x)}\over{dE_x}}dE_e\,. 
\nonumber  
\end{eqnarray} 
It is easy to show that this condition for $\gamma^\prime\geq 1$ reduces to 
\begin{equation} 
{{K^\prime_p}\over{K^\prime_e}}\left({m\over M}\right)^{\gamma^\prime -1}=1\, 
\label{e_p} 
\end {equation} 
on the assumption that the spectral index of electrons and 
protons is $\gamma^\prime$. 
 
\begin{figure}[thbp]  
\mbox{
\includegraphics[width=1.\hsize,clip]{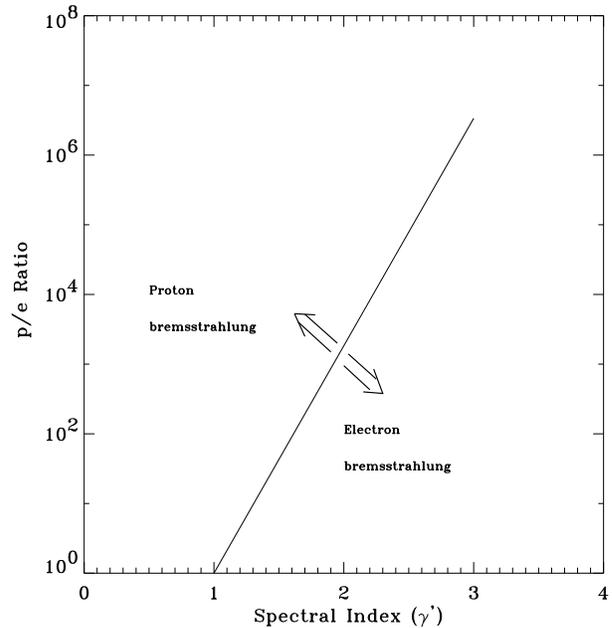}
}    
\caption{The ranges of spectral parameters for predominantly proton or electron bremsstrahlung emission.  
\label{fl_r}}  
\end{figure}

If this ratio is less than unity then the photon flux is mainly 
produced by electrons, and otherwise by protons. The ratio 
$K^\prime_p/ K^\prime_e$ (which is the proton to electron ratio 
at the same energy) determined from Eq.(\ref{e_p}) is shown in 
Fig.\ref{fl_r} as a function of the spectral index 
$\gamma^\prime$.

\section{Processes of Particle Acceleration in the Galactic Ridge} 
 
Below we summarize the characteristics of the  hard X-ray 
spectrum  which are essential for our analysis of the 
acceleration process in the Galactic disk (see Yamasaki et 
al.1997, Valinia and Marshall 1998, Valinia et al. 2000):

\begin{itemize} 
\item  Features of the diffuse emission in the hard X-ray band 
suggest a diffuse and nonthermal origin; 
\item The large scale association of the hard X-ray emission 
with the thermal X-rays  implies that these two components are 
linked. This leads to the idea that thermal particles in the 
hot plasma are accelerated to produce the nonthermal particles 
responsible for the hard X-ray emission; 
\item  The X-ray flux is produced in the region where the 
particles are  freshly accelerated; 
\item The efficiency of the process emitting the hard X-rays 
drops above $40$ keV; 
\item The regions of particle acceleration are supposed to be 
regions of very hot plasma with parameters: density $n\simeq 
8\cdot 10^{-2}$ cm$^{-3}$ and the temperature $T\simeq 2.6$ 
keV. The filling factor of this plasma  is  taken to be 
$\xi\simeq 10^{-3}$.   
\end{itemize} 
 
We see that the data definitely point to  acceleration 
operating in  the Galactic ridge with the emitting particles 
being accelerated from the hot thermal pool. Here we return to 
the old idea of Fermi on CR acceleration in the interstellar 
medium. In the current theory of  cosmic-ray origin this 
mechanism of  interstellar acceleration (or reacceleration) is 
considered as an auxiliary process only slightly changing   the 
spectrum and chemical composition of relativistic CRs emitted 
by SNR (see Berezinskii et al. 1990). It may  follow from the 
X-ray data that the processes of  interstellar acceleration 
play instead the main role in the production of the 
subrelativistic CRs in the inner Galaxy.  
 
The main process of charged particle acceleration in cosmic 
plasmas is in  most cases  collisions with magnetic field 
fluctuations, which leads to slow stochastic energy gain. If this magnetic 
turbulence is `weak', i.e. the amplitude of magnetic field fluctuations $\delta H$ is much less than the strength of the large scale magnetic field $H_0$, 
$\delta H\ll H_0$, then the interaction between the particles and the 
fluctuations has a resonance character. This type of interaction takes place 
in the interstellar medium. The acceleration is more effective for the case of 
strong magnetohydrodynamic turbulence ($\delta H\sim H_0$). This acceleration 
is realized e.g. in low ionized turbulent media (see Dogiel et al. 1987) or in  extended  regions of hot plasma (OB associations) filled 
with numerous but rather weak shock waves, resulting from SN explosions there 
(see Bykov and Toptygin 1993). 
The stochastic acceleration is described in this case as momentum diffusion with the 
diffusion coefficient $\alpha (p)$  
\begin{equation} 
\alpha(p)= \alpha_0 p^2\,, 
\label{m_dif} 
\end{equation} 
where the value $\alpha_0^{-1}$ is the characteristic time of the stochastic 
acceleration.  The production spectrum of the accelerated particles is very hard and it has the form in subrelativistic and relativistic energy ranges  
\begin{equation} 
Q(E)\propto E^{-1}\,. 
\label{stoch} 
\end{equation} 
This circumstance allows in principle to distinguish between stochastic acceleration (\ref{m_dif}) and  acceleration  by supernova shocks which produce softer spectra.  
 
The stochastic acceleration (\ref{m_dif}) will be 
analysed below. 
 
\section{Number of Accelerated Particles} 
 
In order to estimate  how many subrelativistic particles can be produced by 
this mechanism we   solve the equation describing  the spectrum in the 
thermal and nonthermal energy ranges. 
 
The equilibrium (Maxwellian) 
spectrum of background charged particles is produced by  
Coulomb collisions. In an ionized plasma which consists of protons and 
 electrons the energy loss of a test particle with  velocity $v$, 
charge $Z$ and  atomic number $A$ due to these collisions has 
the form (see, e.g., Butler and Buckingham 1962, Sivukhin 1964)  
\begin{eqnarray} 
 \label{de_dt}  
{{dE}\over{dt}}&=-&{{4\pi Z^2 e^4  
n\ln\Lambda}\over{AMv}}\cdot\\  
&&\left(G(v/v^p_T)+(M/m)G(v/v^e_T)\right)\,,  
\nonumber  
\end{eqnarray}  
where  
\begin{eqnarray}  
G(x)= &&{2\over\sqrt{\pi}}\Biggl[\int\limits_0^x  
\exp(-z^2)dz-\\ 
&&\left(1+{\hat m\over M}\right)x\exp(-x^2)\Biggr]\,,  
\nonumber   
\end{eqnarray}  
$\hat m$ is the rest mass of a background charged particle,  $v^p_T$ and $v^e_T$  
are the thermal velocities at the temperature $kT$ of the background protons  
 and electrons  
\begin{equation}  
  v^p_T=\sqrt{{2kT}\over{M}}~~~\mbox{and}~~~ v^e_T=\sqrt{{2kT}\over{m}}\,,  
 \end{equation}  
 and  
\begin{equation}  
 \Lambda={{kTd}\over{e^2}}  
\end{equation}  
Here  $d$ is the Debye radius  
\begin{equation}  
\label{deb}  
d=\sqrt{{kT}\over{8\pi n e^2}}\,.  
\end{equation}  
From Eq.(\ref{de_dt}) it is clear that the  energy loss rate of ions has  at 
least two maxima at 
velocities $v\sim v^p_T$ and  $v\sim v^e_T$.  
 
The equation for stochastic acceleration forming the spectrum of nonthermal 
particles and Coulomb collisions forming an equilibrium Maxwellian spectrum of 
background particles in the energy range  above $kT$  has the form  
\begin{equation}  
{{\partial f}\over{\partial\tau}}-{1\over  
u^2}{\partial\over{\partial u}}\left(A(u){{\partial f}\over{\partial u}}+B(u)  
f\right)=0\,.  
\label{kineq} 
\end{equation}  
Here $f$ is the particle distribution function, $\tau$ and $u$ are the 
dimensionless time and the proton velocity: 
\begin{equation}  
\tau=\nu_0 t~~~\mbox{and}~~~ u={v\over{\sqrt{kT/{\hat m}}}}\,,  
\end{equation}  
$\nu_{0}$ is the  collisional frequency of background particles at $E=kT$ and 
$\hat m$ is the rest mass of the accelerated particles. The frequency $\nu_0$ 
is  
\begin{equation}  
\nu_0={{4\pi nZ^2e^4}\over{(kT)^{3/2}{\hat m}^{1/2}}}\ln\left({{kTd}  
\over{e^2}}\right)\,.  
\label{nup}  
\end{equation}

To estimate the number of  accelerated particles we must analyze the range of 
velocities $u>1$  where the functions $A(u)$ and $B(u)$  are 
 \begin{equation}  
A(u)={1\over  
u}+{1\over u}{M\over m}G\left({u\sqrt{m\over {2M}}}\right)+\alpha_0^p(u) u^4  
\end{equation}  
and  
\begin{equation}  
B(u)=1+{M\over m}G\left({u\sqrt{m\over {2M}}}\right) 
\end{equation}  
for protons and 
 \begin{equation} A(u)={1\over  
u}+\alpha_0^e(u) u^4  
\end{equation}  
and  
\begin{equation}  
B(u)=1 
\end{equation} 
for electrons. 
  
The nondimensional momentum diffusion coefficient $\alpha(u)$ describes 
processes of stochastic acceleration which in many cases has the form (see, 
e.g., Toptygin 1985)  
\begin{equation} 
\alpha(u)=\alpha_0 u^2\,,  
\end{equation} 
and  the other terms describe Coulomb collisions with the background  protons 
and electrons. For the same acceleration frequency $\alpha_0^\prime$ the values 
of the nondimensional $\alpha_0$ are different for  electrons and protons: 
$\alpha_0^{e,p} = \alpha_0^\prime/\nu_0^{e,p}$. 
 
\begin{figure}[thbp]  
\mbox{
\includegraphics[width=1.\hsize,clip]{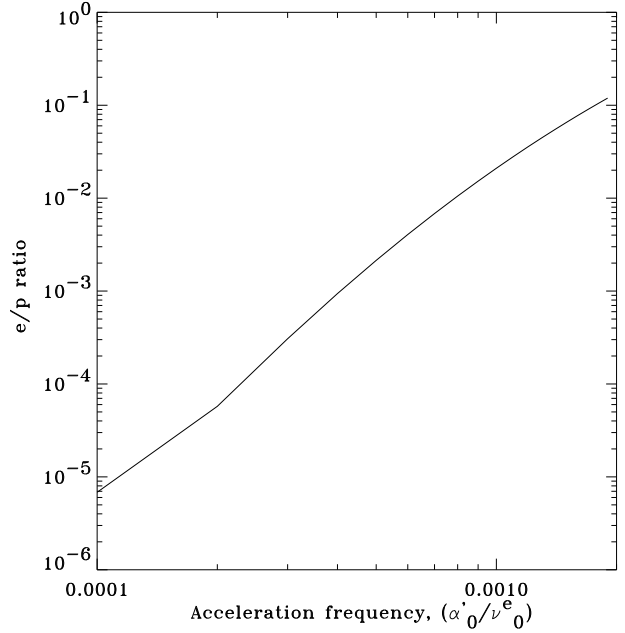}
} 
 
\caption{The ratio of the run-away fluxes (electrons to protons). The 
dimensionless acceleration frequency $\alpha_0$ is normalized to the collision 
frequency $\nu^0_e$.  
\label{spe}}  
\end{figure}

This acceleration violates the equilibrium state of the particle distribution 
determined  by Coulomb collisions since  particles with energies above the 
injection energy (where acceleration is significant)  continuously increase 
their energy. Collisions tend to compensate this particle loss causing the 
energy of thermal particles to increase from the main equilibrium value to the 
injection energy and as a result a flux of  particles `running away' into the 
high energy range occurs in the spectrum even at energies far below the 
injection energy.   The value of the total  run-away flux (for protons and 
electrons) has the form (see Gurevich 1960) 
\begin{eqnarray} 
\label{pac} 
&&\left({{d{\bar N}}\over{dt}}\right)_p^r\simeq\sqrt{2\over\pi}n 
\nu_0^p(1+M/m)\cdot\\ 
&&\exp\left(-{\int\limits_0^\infty}{{udu(1+(M/m)G(u/\sqrt{2M/m}))}\over{1+(M/m)G(u/\sqrt{2M/m})+\alpha_0^p u^5}}\right)\,, 
\nonumber 
\end{eqnarray} 
\begin{equation} 
\label{eac} 
\left({{d{\bar N}}\over{dt}}\right)_e^r\simeq\sqrt{2\over\pi}n 
\nu_0^e 
\exp\left(-{\int\limits_0^\infty}{{udu}\over{1+\alpha_0^e u^5}}\right)\,. 
\nonumber 
\end{equation} 
The ratio of the proton and electron run-away fluxes for different values of 
the nondimensional parameter $\alpha_0^e=\alpha^\prime_0/\nu_0^e$ are shown in 
Fig.\ref{spe}. 
 
From Fig.\ref{spe} we see that protons are accelerated much more effectively 
than electrons. This means that in the case of interstellar acceleration it is 
easier to produce nonthermal protons than electrons. 
 
As  follows from the analysis of the kinetic equation (\ref{kineq}) the 
spectrum of  accelerated particles is hard and and has the form $E^{-1}$. The 
production rate  $Q$ of  particles can be determined from the value of their 
run-away flux as 
\begin{equation} 
Q(E)=K\cdot E^{-1}\theta (E_{max}-E)\,, 
\label{q_s} 
\end{equation} 
where Eqs. (\ref{pac}) and (\ref{eac}) determine the constants $K$ for protons 
and electrons, and $\theta (x)$ is  the Heaviside  function (step-function). 
The energy $E_{max}$ determines a cutoff in the spectrum  where the efficiency 
of acceleration drops.

 For  stochastic acceleration  $E_{max}$ can be derived from the spectrum of electromagnetic turbulence  (see e.g. Dogiel et al. 1987). However, since we do not specify the mechanism of acceleration its value is uncertain though e.g.  for the case of supersonic turbulence  in OB associations the maximum energy of accelerated protons was estimated  to be   of the order of 100 MeV (see  Bykov and Bloemen, 1997). The maximum energy of electrons can be smaller than hundreds of keV (see section 10). 

 In general the value of $E_{max}$ is determined by the efficiency of particle acceleration and by processes which prevent particle acceleration above a certain energy (e.g. particle escape from the acceleration region or the feedback reaction of accelerated particles on the efficiency of acceleration). As we mentioned already we do not know the details of the acceleration processes in the galactic disk, and therefore we are unable to estimate the value of $E_{max}$ in this way.

Fortunately, in our case the value of $E_{max}$ can be directly derived from the observed X-ray spectrum. 
To reproduce the X-ray spectrum (\ref{sp_val}) we must assume that the maximum 
energy  does not exceed significantly the value of $100$ keV for the 
accelerated electrons or $100$ MeV for the accelerated protons, which is close 
to the value  from theoretical estimates for the interstellar acceleration of 
nuclei  (see, e.g., Bloemen and Bykov 1997).

\section{Chemical Composition of the Accelerated Particles} 
The  interstellar stochastic acceleration is described as momentum diffusion 
with the diffusion coefficient 
\begin{equation} 
\alpha(p)=\alpha_0^\prime p^2\,. 
\end{equation} 
Then the average rate of acceleration can be determined as 
\begin{equation} 
{{dE}\over{dt}}\simeq \alpha_0^\prime E\,. 
\end{equation} 
 
It is clear that the acceleration generates a particle spectrum at energies 
higher than the injection energy which is determined from the equality 
between the ionization losses and the acceleration 
\begin{equation} 
 E_{inj}\simeq Z^{4/3}A^{1/3}\left({{4\pi e^4 n \ln\Lambda\sqrt{M_p}} 
\over{\sqrt{\pi} m \alpha_0^\prime}}\right)^{2/3}\,, 
\label{einj} 
\end{equation} 
where $A$ is the particle atomic number. We see that the flux of the 
accelerated particles is a function of the particle charge $Z$. Therefore it 
would not be surprising if the chemical composition of the accelerated 
particles differs from the chemical composition of the background plasma. 
 
The flux of ions escaping from the thermal part of the spectrum into the 
acceleration region  can be represented as (Gurevich 1960) 
\begin{eqnarray} 
\label{cc} 
{{df}\over{dt}}=&&\eta (Z)\sqrt{2\over \pi}n {Z^2\over A^{0.5}}\nu_0^p {M\over m} \cdot\\ 
&&\exp\left(-{{Z^2q}\over {2\alpha_0^p A^{0.5}}}\sqrt{m\over M}\right)\,, 
\nonumber 
\end{eqnarray}    
where $q$ is of the order of unity and $\eta(Z)$ is the abundance in the 
background gas. From this equation we see that the abundance  of ions with 
$Z>1$ in the flux of accelerated particles is strongly suppressed compared to 
the abundance of elements in the background plasma and therefore the abundances 
of the background gas and that of the accelerated nuclei are quite different. 
 
\begin{figure}[thbp]  
\mbox{
\includegraphics[width=1.\hsize,clip]{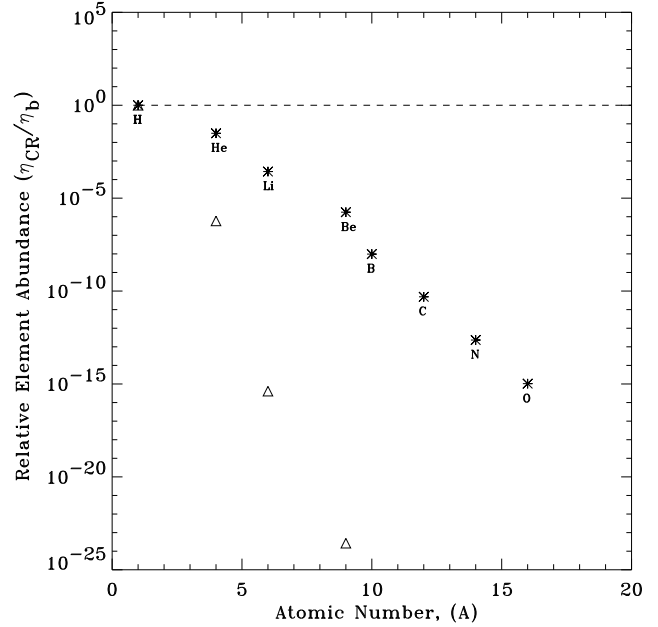}
} 
 \caption {The relative abundance of elements in the background gas (straight 
dashed line) and  in the flux of particles ($\eta_{CR}/\eta_b$) accelerated 
from the thermal pool (asterisks), calculated for $\alpha_0^\prime/\nu_0^p 
=10^{-2}$. The result of calculations for the acceleration parameter 
$\alpha^\prime_0/\nu_o^p=8.85\cdot 10^{-4}$ is shown by triangles.  
\label{abund}}  
\end{figure}

The chemical composition of subrelativistic nuclei accelerated from the thermal 
pool is shown in Fig.\ref{abund}. In the figure we show the ratio of the 
nuclear abundance in the flux of the accelerated particles $\eta_{CR}$ to the 
abundance in the background gas $\eta_b$.  The abundance of element in the 
background gas is taken as unity for all elements (straight dashed line). The 
chemical composition of the accelerated particles was calculated for 
acceleration frequencies $\alpha_0^\prime/\nu_0^p=10^{-2}$ shown by asterisks and  for 
$\alpha^\prime_0/\nu_o^p=8.85\cdot 10^{-4}$  shown by triangles.  For the 
calculations we used Eq.(\ref{pac}) which gives a more accurate result than the approximation (\ref{cc}).  

\section{Spatial Distribution of the Subrelativistic Particles   in the Galaxy} 
The transformation of the injection spectrum (\ref{q_s}) in the interstellar 
medium is due to processes of energy losses and spatial propagation. The cosmic 
ray propagation is described as diffusion in the interstellar medium with the 
effective diffusion coefficient $D$ and the convection velocity $u$. The 
propagation equation in the general form is 
\begin{equation} 
\nabla ({u}N-D\nabla N)-{{\partial}\over {\partial E}} \left({{dE}\over{dt}}N\right)=Q(r,E)\,, 
\label{diffu} 
\end{equation} 
where $dE/dt$ describes the rate of energy losses. 
 
Below we obtain rather simple solutions of the one dimensional propagation 
equation for  subrelativistic protons and electrons. The particle production 
for the case of the interstellar acceleration is described by Eq.(\ref{q_s}). 
The spectrum of particles  depends strongly on the particle mean free path 
$\lambda$ determined by the processes of diffusion and energy losses  
\begin{equation} 
\lambda\sim\sqrt{D \tau_l}\,, 
\end{equation} 
where $\tau_l$ is the characteristic life-time of the particles determined by 
the energy losses. There are  three other spatial scales which characterize the 
distribution of the CR sources $z_s$, the gas $z_g$ and the volume of the 
particle propagation (halo) $z_h$. 
 
If $\lambda<z_s$ the propagation is unimportant and the kinetic equation has 
the form 
\begin{equation} 
{d\over {dE}}\left({{dE}\over{dt}} N\right)=KE^{-\gamma}\,, 
\label{leaky} 
\end{equation} 
where the right hand side of the equation describes the injection spectrum and $dE/dt\simeq -a/E^{0.5}$ is the rate of ionization losses. The solution of this equation is 
\begin{equation} 
N={{KE^{-(\gamma -3/2)}}\over{\mid \gamma-3/2 \mid a}}\,. 
 \end{equation} 
 
If $z_z<\lambda<z_g$ particles escape from the acceleration region and the 
equation has the form 
\begin{equation} 
D{{\partial^2} \over {\partial {z^2}}}{N}+{\partial \over {\partial  
E}}\left({{dE}\over{dt}}N\right)=-K{E^{-\gamma}}{z_s}\delta (z) 
\end{equation} 
and the solution is 
\begin{equation} 
N(E)={{Kz_s}\over{\mid dE/dt\mid}}{\int\limits_0^\infty} dE_0\exp\left(-{{z^2}\over{4D\tau}}\right) 
{{E_o^{-\gamma}}\over\sqrt{4D\tau}} 
\end{equation} 
where  
\begin{equation} 
\tau(E, E_0)={\int\limits_{E_0}^E}{{dE^\prime}\over{dE^\prime/dt}}\,. 
\end{equation}

If the particles lose their energy in the gaseous disk ($\lambda<z_g$) then we 
obtain 
\begin{equation} 
{N}\sim K \sqrt{{a{t_d}}\over {4\pi}} {E^{-(\gamma-{3\over 4})}} 
\end{equation} 
and if particles leave the gaseous disk, then the energy losses are unimportant 
and the solution is 
\begin{equation} 
N\simeq {{KE^{-\gamma}z_h^2}\over D}\left(1-{z\over z_h}\right)\,. 
\end{equation} 
 
We see that the effect of  ionization energy losses is a flattening of the 
accelerated spectrum. Therefore we expect that the spectral index 
$\gamma^\prime$ of particles in the interstellar medium ($N\propto 
E^{-\gamma^\prime}$) is equal to or less than the injection spectral index 
$\gamma$ ($Q\propto E^{-\gamma}$). The injection index $\gamma$ of subrelativistic 
particles accelerated by shock waves is $1.5$ and that for the stochastic 
acceleration by chaotic electromagnetic turbulence is $1$. Therefore we expect 
that the spectral index of fast particles in the interstellar medium is in the 
range $-0.5\leq\gamma^\prime\leq 1.5$. 
 
\section{Bremsstrahlung X-Ray Emission from the Ridge} 
From Sects. 4 and 6 we concluded that the spectrum of the X-ray emitting 
particles is a power-law with an exponential cut-off 
\begin{equation} 
N(E)\simeq K(\gamma^\prime) E^{- \gamma^\prime}\exp\left(-{E\over{E_{max}(\gamma^\prime)}}\right)\,. 
\end{equation} 
Then we can formally derive the values of $K(\gamma^\prime)$ and $E_{max}$ for 
different values of $\gamma^\prime$  from the observed intensity of the hard 
X-ray emission using the equation for bremsstrahlung radiation (\ref{brf}).

\begin{figure}[thbp]  
\mbox{
\includegraphics[width=1.\hsize,clip]{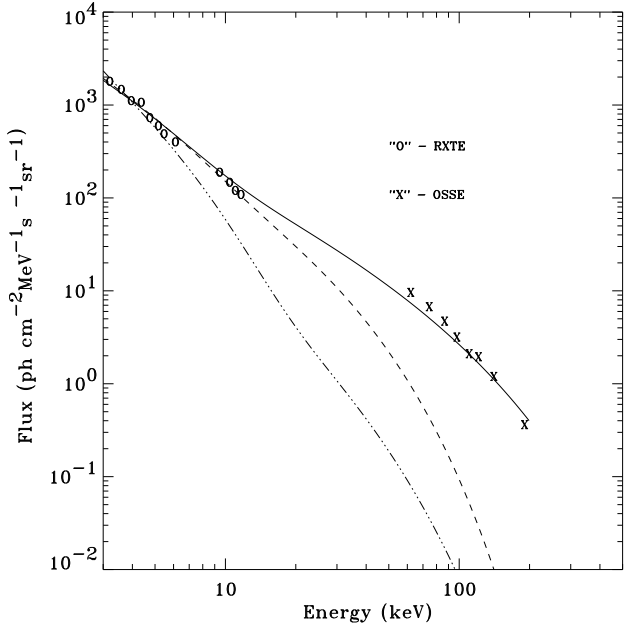}
} 
\caption{The flux of hard X-ray emission measured by the RXTE and OSSE 
telescopes. The  spectrum of bremsstrahlung radiation (generated by protons or 
electrons) reproducing the RXTE-OSSE Spectrum I (solid line) and that for the 
RXTE Spectrum II only (dashed line).  The spectrum for 
$\gamma^\prime=2$ is shown by the dashed-dotted line. 
The parameters 
$E_{max}(\gamma^\prime$) and $K(\gamma^\prime$) were derived for each value 
of the spectral index $\gamma$ to satisfy the observational data. 
\label{vl}}  
\end{figure}

Below we denote as Spectrum I the spectrum of the emitting particles which 
produce the hard X-ray emission in the range $10-200$ keV shown in Fig.\ref{vl} by the solid line. The spectrum of the particles which produce the emission in the GINGA-RXTE range only (shown by the dashed line in Fig.\ref{vl}) will be 
denoted as Spectrum II.

We took the gas column density in the direction of the Galactic ridge as $\sim 
10^{22}$ cm$^{-2}$ assuming that the accelerated particles fill the whole 
volume of the ridge. 
 
\begin{figure}[thbp]  
\mbox{
\includegraphics[width=1.\hsize,clip]{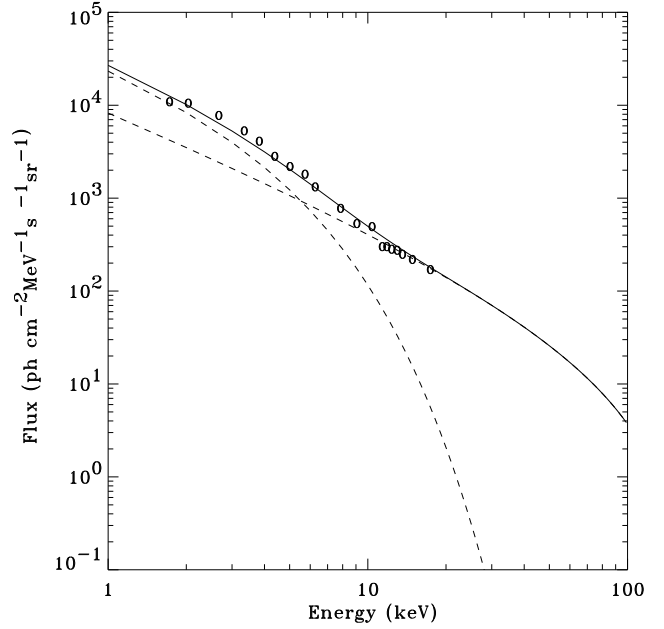}
} 
\caption{The flux of hard X-ray emission measured by the GINGA telescope for the central Galactic radian. The  bremsstrahlung radiation was computed for a spectral index $\gamma^\prime=0.1$. Spectra of the thermal and nonthermal emission are also show in the figure. 
\label{ym}}  
\end{figure}

The spectra of hard X-rays  from GINGA presented in Yamasaki et al. (1997) and 
  from RXTE by Valinia and Marshall (1998) were observed for different regions 
  of the Galactic ridge and therefore their intensities differ  by a factor of 
  3 -- 5. Hence  we derived the spectrum separately for these observations. The  results of the measurements are presented in Figs. \ref{vl} and \ref{ym}. 
 The densities of the particles derived from the RXTE and GINGA data  differ 
 also  by a factor  3 -- 5.  To present the results in the same units we took 
 the total area of the central radian observed by GINGA, namely $456$ grad$^2$ 
 (the average  width of the ridge is $\sim 8^o$   (Valinia, private 
 communication). Below we base our analysis on the RXTE data since they 
 represent the average X-ray spectrum in the Galactic ridge.  In both cases we 
took the component below $10$ keV to be due to the thermal bremsstrahlung at a 
temperature of $2.6$ keV.

The results of our calculations for the RXTE data are shown in Fig. \ref{vl}. 
 We obtained almost the same X-ray spectrum for different $\gamma^\prime$ and 
 $E_{max}(\gamma^\prime)$   by adjusting these parameters.

 As an example we show in Table 1 for Spectrum I and Spectrum II how the break position $E_{max}$ in the spectrum of protons changes with the value of the spectral index $\gamma^\prime$. 

\begin{table} 
\caption{Parameters of the emitting spectrum of protons (p) consistent with for Spectrum I and Spectrum II. For electrons the position of the break can be obtained from $E_{max}^e=E_{max}(m/M)$} 
\begin{tabular}{|p{1.cm}|p{1.cm}|p{1.cm}|p{1.cm}|p{1.cm}|} 
\hline  
$\gamma^\prime$&1&0.5&0.1&-0.5\\
\hline
$E_{max}^I$&470&300&250&150\\
MeV&&&&\\ 
\hline
$E_{max}^{II}$&65&55&45&35\\
MeV&&&&\\
\hline
\end{tabular}
\end{table}  

 Spectra for all 
values of $\gamma^\prime \leq 1$ nicely describe the RXTE data. It is clear 
that steeper electron spectra  are unable to reproduce the   flat RXTE 
spectrum. The result of calculations for $\gamma^\prime=2$ are shown in the 
figure.  From  Eq.(\ref{brf}) the X-ray spectrum is then proportional to 
$E_x^{-2.5}$ which is much steeper than the observed spectrum at energies below $30$ keV. 
 
Similar calculations based on the GINGA data are shown in Fig. \ref{ym}. 
Qualitatively the conclusion is the same - the ridge emission is produced by a 
flat spectrum of emitting particles whose spectral index is 
$\gamma^\prime\le 1$.

In principle we can extend our   calculations  to the region of the OSSE data 
up to $200$ keV in order to reproduce the total hard X-ray flux from the ridge 
by the bremsstrahlung emission of the accelerated particles. However as we 
remarked in  the Introduction the origin of this part of the ridge spectrum is 
unclear. Results of the calculations are shown in Fig.\ref{vl} by the solid 
line. In the range between $200$ to $500$ keV part of the flux is from 
three-photon positronium continuum emission. Therefore, OSSE data in this 
energy range are excluded in Fig.\ref{vl}.    

The important conclusion  following from these calculations is: the data can be reproduced by  bremsstrahlung emission of the accelerated particles only if 
$\gamma^\prime\leq 1$ . The cut-off energy $E_{max}$ for Spectrum I is of the 
order of $150-500$ MeV for protons and $80-270$ keV for electrons.  The cut-off energy for Spectrum II equals $35-50$ MeV for protons and $20-30$ keV for 
electrons.

Below we consider separately the details of    protons and electrons as the 
origin of the hard X-rays.

\section{Proton Bremsstrahlung Origin of the Ridge X-Rays} 
\subsection{Parameters of the proton spectrum} 
First of all we estimate  from the ridge X-ray spectrum  the total density of 
protons  $\bar N$  
\begin{equation} 
\bar N=\int\limits^{\infty}N(E)dE\,, 
\end{equation} 
their energy density $w$ 
 \begin{equation} 
w=\int\limits^{\infty}EN(E)dE\,, 
\end{equation} 
the total energy power $F$ (if the particle loss from the ridge is due to the 
ionization losses (\ref{il})) 
\begin{equation} 
F=V\int\limits^{\infty}N(E){{dE}\over{dt}}dE\,, 
\end{equation} 
where  $V$ is the volume of the emitting region of $10^{67}$ cm$^3$ (Valinia 
and Marshal 1998), and the total (integrated over energies) particle flux $d{\bar N}/dt$ due to the losses which 
has to be compensated by the acceleration 
\begin{equation} 
\left({{d{\bar N}}\over{dt}}\right)_i\sim\int\limits^{\infty}{{N(E)}\over \tau_i}dE\,, 
\label{i} 
\end{equation} 
with $\tau_i$ from Eq.(\ref{taui}). 
 
The results of the calculations for the emitting protons on the assumption that 
they fill the whole Galactic disk (filling factor $\xi=1$) are presented in 
Table 2. 
\begin{table} 
\caption{Parameters of the accelerated protons (p) and electrons (e)} 
\begin{tabular}{|p{0.1cm}|p{1.07cm}|p{1.6cm}|p{1.45cm}|p{1.23cm}|p{1.3cm}|} 
\hline 
&${\bar N}$&$w$&$d{\bar N}/dt$&$\alpha_0^\prime$ &$F$\\ 
&(cm$^{-3}$)&(eV cm$^{-3}$)&(cm$^{-3}$s$^{-1}$)&(s$^{-1}$) &(erg s$^{-1}$)\\ 
\hline 
&&&&&\\ 
p&$2\cdot 10^{-6}$&$250-400$&$\sim 10^{-21}$&$7\cdot 
10^{-15}$&$\sim 10^{42}$\\ 
&&&&&\\ 
\hline 
\hline 
&&&&&\\ 
e&$2\cdot 10^{-6}$&$0.2$&$\sim 10^{-18}$&$3\cdot 10^{-13}$&$\sim10^{42}$\\ 
&&&&&\\ 
\hline 
\end{tabular} 
\end{table} 
These parameters calculated for Spectrum I and Spectrum II do not differ 
greatly.  
 
\begin{figure}[thbp]  
\mbox{
\includegraphics[width=1.\hsize,clip]{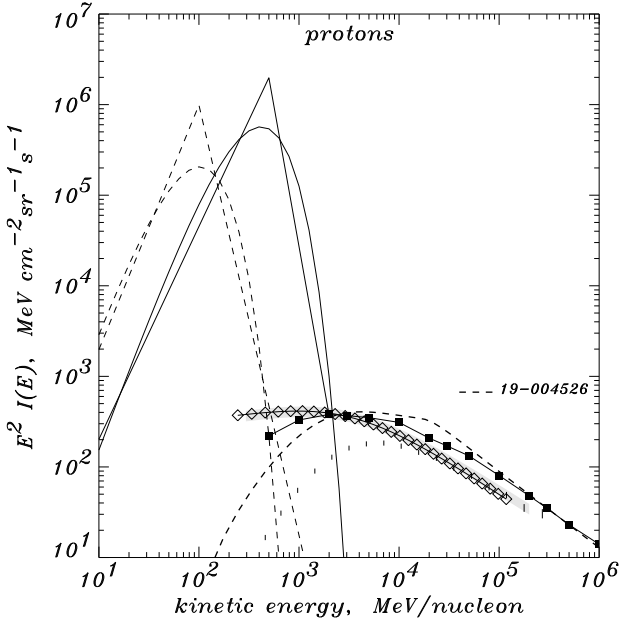}
}  
 \caption{The function $E^2\cdot I(E)$ for the observed relativistic and derived subrelativistic spectra of the interstellar protons (solid line - Spectrum I; thin dashed line - Spectrum II), vertical bars -  IMAX direct measured spectrum, thick dashed line the proton spectrum derived from the propagation model, connected squares and diamonds, and dashed area - different variants of the proton demodulated spectrum (see Strong et al. 2000 and references therein). We show also in the figure  (triangles)  the approximations to  Spectrum I and Spectrum II which were used for estimates of the $\pi^o$ gamma-ray flux produced by the accelerated protons. 
\label{pr_sp}}  
\end{figure}

We see that the energy density of the subrelativistic protons must be very high in order to reproduce the X-ray data. In Fig. \ref{pr_sp} we present the 
expected Spectrum I (solid line) and Spectrum II (dashed line) for the 
subrelativistic protons together with the  spectrum of  protons  at Earth .  
 
The equation for $\alpha_0$ can be derived from the balance between the 
run-away flux of the accelerated particles and the nonthermal particle losses due to 
ionization. This balance fixes the  density of the accelerated particles in the disk.  
The total production rate of the accelerated particles  is determined by the momentum diffusion with the coefficient $\alpha_0$. 
Therefore the equation for $\alpha_0$ (which follows from Eq.(\ref{diffu})) has the form 
\begin{equation} 
\left({{d\bar N}\over{dt}}\right)_i\simeq \left({{d{\bar N}}\over{dt}}\right)^r 
\label{est} 
\end{equation} 
Here the LHS can be estimated from Eq.(\ref{i}) where the density of the emitting particles is derived from the observed flux of hard X-rays. The RHS of this equation is determined by Eqs.(\ref{pac}) and (\ref{eac}) for protons and electrons respectively, and  clearly it is a function of the parameter $\alpha_0$ only.  

 The flux of  run-away particles $\left({{d\bar N}/{dt}}\right)^r$ is formed by collisions of thermal particles whose distribution is not sensitive to the nonthermal part of the spectrum. Therefore, as one can see from Eqs.(\ref{pac}) and (\ref{eac}) this run-away flux is a function of the parameter $\alpha_0$ only. 

The LHS of Eq.(\ref{est}) is completely determined by the characteristics of the observed spectrum of hard X-ray emission from the Galactic Ridge. As  was shown in Section 8 the maximum energy of emitting particles chosen to reproduce the spectrum of X-rays can vary over rather wide limits (depending on assumption of the contribution from unresolved sources). However, these variations only weakly (logarithmically) change the derived value of $\alpha_0$. On the other hand, the density of accelerated particles strongly (exponentially) depends on $\alpha_0$ and even small variations of $\alpha_0$ change this density significantly.

From Eq.(\ref{est}) we find that  the proton acceleration 
frequency equals $\alpha_0^\prime/\nu_0^p\simeq 1.3\cdot 10^{-3}$ almost independent of the $\gamma^\prime$ value. 
 
The  frequency of proton collisions 
$\nu_0^{p}$    at energy $E=kT$ (see Eq.(\ref{nup})) in the acceleration 
regions is: 
\begin{equation} 
\nu_0^p\simeq 5.3\cdot 10^{-12}~\mbox{sec$^{-1}$}\,. 
\end{equation} 
Here we took $n\simeq 8\cdot 10^{-2}$ cm$^{-3}$ and $T\simeq 2.6$ keV for regions of the interstellar medium heated by SNs (see Yamasaki et al. 1997).   
 
Then the characteristic acceleration frequancy is 
\begin{equation} 
 \alpha_0^\prime \sim 7.1\cdot 10^{-15}~ \mbox{sec$^{-1}$} 
\label{apd} 
\end{equation} 
for protons. 
 
\subsection{Gamma-Ray Line Emission from the Galactic Ridge} 
 
The chemical composition of the accelerated ions is important for estimates of 
gamma-ray  line emission from the Galactic ridge which is expected in the case 
of a nuclear bremsstrahlung origin of the X-ray emission. The most prominent 
4.4 MeV \carbon and 6.1 MeV \oxygen line flux from the Galactic ridge was 
estimated by Pohl (1998) and Valinia et al. (2000b) for hypothetical spectra 
of subrelativistic protons and electrons. Their conclusions were not in favor 
of the inverse bremsstrahlung model since the calculated line flux was much 
higher than the upper limits derived from the OSSE and COMPTEL observations. On the other hand they used for their estimates the observed CR chemical 
composition which can differ from the chemical composition of subrelativistic 
nuclei accelerated from the thermal pool. Hence we re-calculated the line flux 
from the ridge in the framework of our model.  
\begin{figure}[thbp]  
\mbox{
\includegraphics[width=1.\hsize,clip]{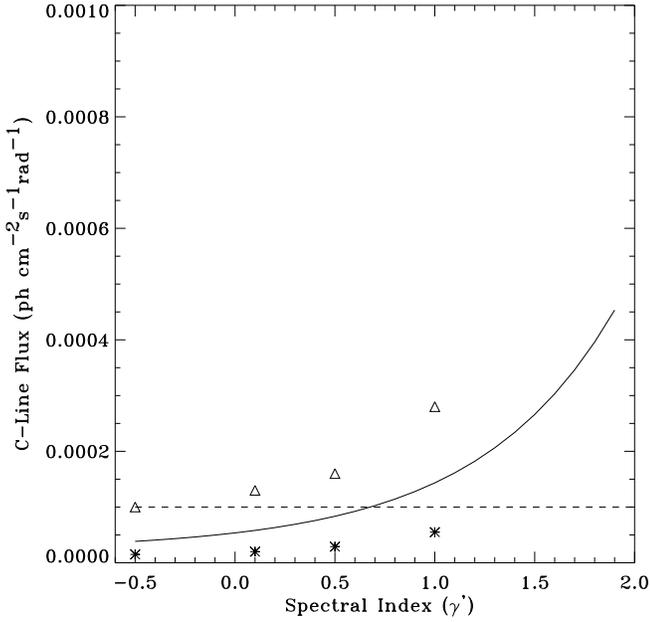}
} 
  \caption {The expected flux in the 4.4 MeV \carbon line calculated for a proton 
spectrum consistent with the RXTE data (solid line)  based on the chemical 
composition of the accelerated nuclei (\ref{cc}).  The straight dashed line 
shows the marginal $3$ to $7$ MeV flux observed with COMPTEL in the direction 
of the Galactic ridge.  The line flux calculated for different $\gamma^\prime$ 
in  Spectrum I is shown by asterisks and that for  Spectrum II  by 
triangles. The parameters 
$E_{max}(\gamma^\prime$) and $K(\gamma^\prime$) were derived for each value 
of the spectral index $\gamma^\prime$ to satisfy the observational X-ray data. 
}  
\label{c_l}  
\end{figure}  
\begin{figure}[thbp]  
\mbox{
\includegraphics[width=1.\hsize,clip]{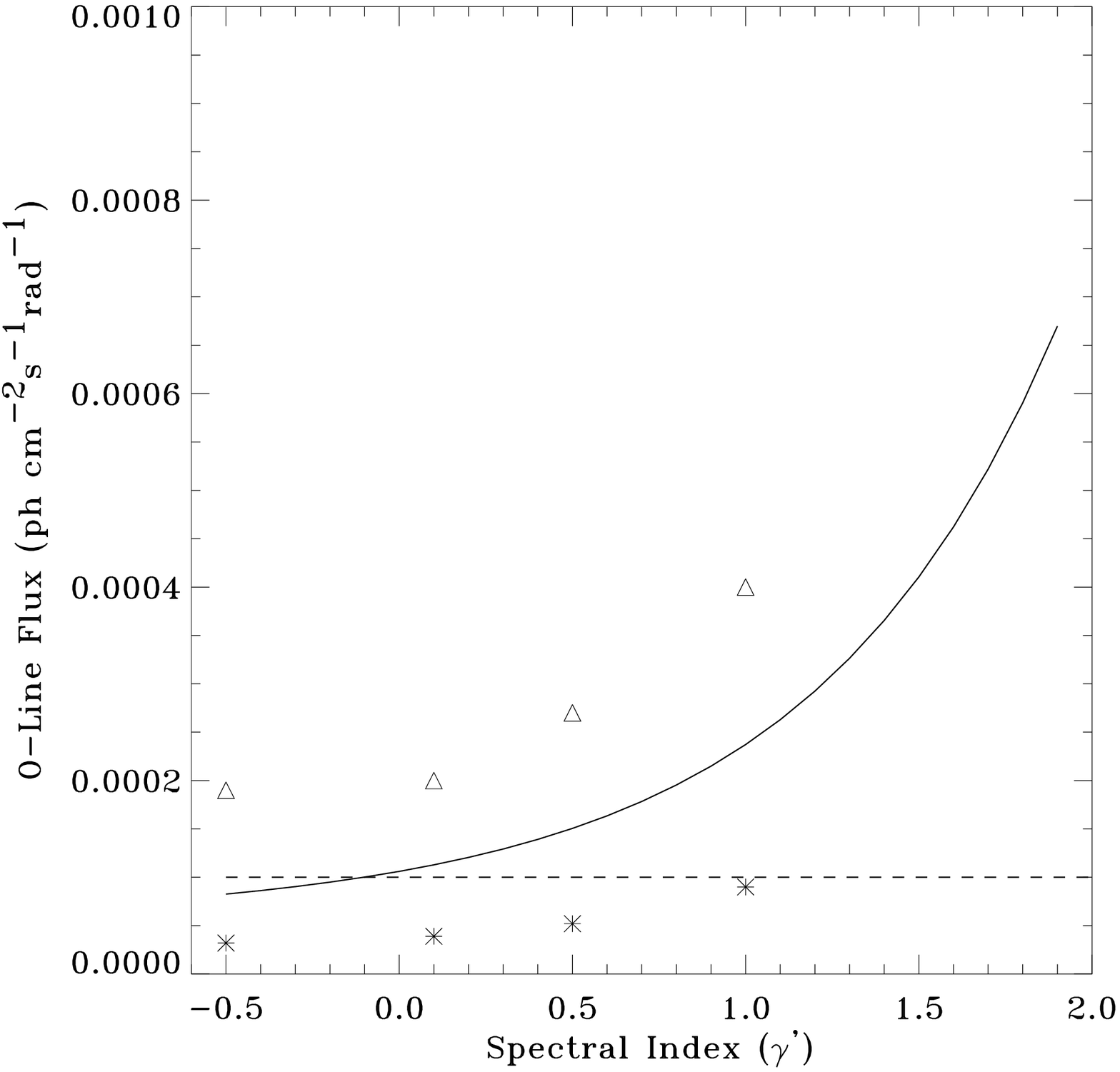}
} 
\caption {The expected flux in the  6.1 MeV \oxygen line calculated for a 
proton spectrum consistent with  the RXTE data (solid line) based on the 
chemical composition of the accelerated nuclei (\ref{cc}). The straight dashed 
line shows the marginal $3$ to $7$ MeV flux observed with COMPTEL in the 
direction of the Galactic ridge. The line flux calculated   for different 
$\gamma^\prime$ in  Spectrum I is shown by asterisks and that for the Spectrum 
II  by triangles. 
The parameters $E_{max}(\gamma^\prime$) and $K(\gamma^\prime$) were derived for  each value 
of the spectral index $\gamma^\prime$ to satisfy the observational X-ray data.}  \label{o_l}  
\end{figure}

The line flux can be estimated from 
\begin{equation} 
Q_\gamma=\int\limits_E \sigma_\gamma \eta n v N_pdE\,, 
\end{equation} 
where $\eta$ is the abundance of \carbon and  \oxygen in the background gas, 
and  $\sigma_\gamma$ is the cross-section for the line production.

For our estimates we  used the solar abundances $\eta_C\simeq 3.63\cdot 
10^{-4}$ and $\eta_O \simeq 8.5\cdot 10^{-4}$ as the  \oxygen and \carbon 
abundances of the background gas. The cross-sections were taken from Ramaty et 
al. (1979). We remind the reader again that in the framework of this model the 
lines are produced in interactions of fast protons with background \carbon and 
\oxygen nuclei only.

The total flux of X-ray emission produced by protons can be written in the 
form 
\begin{equation} 
Q_X={\int\limits_{(M/m)E_x}}\sigma_{X}n v N_pdE\,, 
\end{equation} 
where the cross-section $\sigma_X$ is taken from Eq.(\ref{sbr}).   Since the 
X-ray and $\gamma$-ray fluxes are produced by the same subrelativistic nuclei 
interacting with the background gas, their ratio is independent of the gas 
density, and for the power-law spectrum of the nuclei (below the cut-off energy 
$E_{max}$) 
\begin{equation} 
N_p=K^\prime (\gamma^\prime)\cdot E^{-\gamma^\prime} 
\end{equation} 
this ratio is  also independent of the density of the subrelativistic cosmic 
rays. If we use as the normalization level the X-ray flux derived from the 
observations   then the expected intensity of the gamma-ray line is 
\begin{equation} 
Q_\gamma=Q_X\cdot{{ \int\limits_E^{E_{max}} \sigma_\gamma \eta  v E^{-\gamma^\prime}dE}\over{ {\int\limits_{(M/m)E_x}^{E_{max}}}\sigma_{X} v E^{-\gamma^\prime}dE }}\,. 
\label{fg} 
\end{equation} 
From Eq.(\ref{fg}) we see that the value of $Q_\gamma$ is a function of the 
unknown spectral index $\gamma^\prime$ only.  
 
The expected flux of the \carbon line from the Galactic ridge calculated based 
on  the RXTE  X-ray flux is shown in Fig.\ref{c_l} and the flux of the  \oxygen 
line is shown in Fig.\ref{o_l}. From these figures we can conclude that the 
limits following from the estimates of the gamma-ray line emission are not very 
restrictive for hard spectra. We notice here that a marginal detection of a $3$ to $7$ MeV excess at the level $\sim 10^{-4}$ ph cm$^{-2}$s$^{-1}$rad$^{-1}$ in the direction of the Galactic center was obtained with the COMPTEL telescope 
(Bloemen and Bykov 1997) which is just  near the level of our estimates. 
 
In  Figs. \ref{c_l} and \ref{o_l} we present also the fluxes of the \carbon and \oxygen line obtained for different parameters ($\gamma^\prime$, 
$E_{max}(\gamma^\prime)$ and $K(\gamma^\prime)$) of the proton Spectrum I 
(asterisks) and Spectrum II (triangles). We see  the position of the 
exponential cut-off is strongly constrained by the upper limit  derived from 
the COMPTEL (and OSSE) data. In the case of rather small values of $E_{max}$ 
($\sim 30-50$ MeV) the line flux generated by the protons exceeds this upper 
limit since in this case the protons are concentrated near the energy where the cross-section for the line production is maximum.  
 
\subsection{The flux of \pio gamma-ray emission produced by the accelerated 
protons} 
\begin{figure}[thbp]  
\mbox{
\includegraphics[width=1.\hsize,clip]{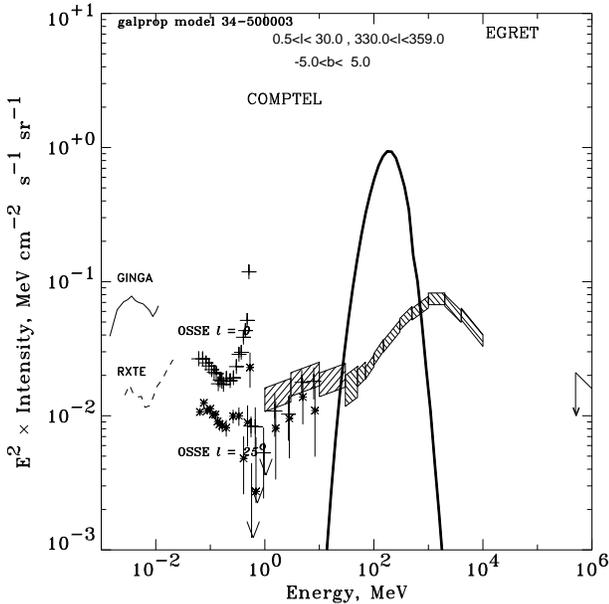}
}  
\caption {The  $\pi^o$  gamma-ray emission produced by  proton Spectrum I.} 
\label{gamma} 
\end{figure}

The \pio gamma-ray photons are generated in collisions of the accelerated 
protons with the background gas.   The flux  was calculated with the model 
described in Strong et al. (2000) and  is shown in Fig.\ref{gamma}.   The 
emission is concentrated in an narrow energy range - between 30 and 1000 
MeV. The lower boundary is determined by the threshold energy of $\pi^o$ 
production and the upper boundary  by the  cut-off in the proton spectrum.  
 
\begin{figure}[thbp]  
\mbox{
\includegraphics[width=1.\hsize,clip]{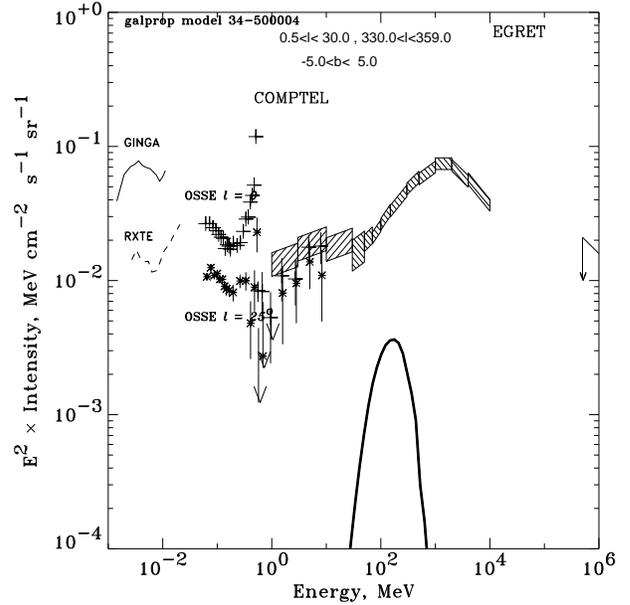}
}   
\caption {The  $\pi^o$  gamma-ray emission produced by  proton Spectrum 
II.} 
\label{gamma_low} 
\end{figure}

The gamma-ray  flux  generated by  protons with Spectrum I is   a factor $100$ 
higher than  measured. The reason is that the number of protons above the 
$\pi^o$ threshold is high  (see  the solid line in Fig.\ref{pr_sp}). 
 
The gamma-ray flux produced by protons with Spectrum II (see 
Fig.\ref{gamma_low}) is negligible when compared with the observed ridge flux 
because of the rather low cut-off energy (see  the dashed line in 
Fig.\ref{pr_sp}). 
 
Hence we conclude that it is problematic to explain the whole nonthermal 
spectrum up to  $200$ keV by  proton bremsstrahlung. However, this model 
successfully  explains the emission up to several tens of keV. In the latter 
case there must be another source of hard X-ray emission at energies $\sim 100$ keV, e.g. unresolved point-like sources.  
It is interesting to notice that a 
similar problem arises in the interpretation of the diffuse Galactic gamma-ray 
data at energies below $30$ MeV where an excess of the emission was found which cannot be produced by bremsstrahlung of the CR electrons and it which may also be due to the emission of unresolved sources (Strong et al.2000). 
 
Thus we see that in addition to the analysis of the gamma-ray line flux  
observed continuum, gamma-ray flux  restricts the value of $E_{max}$ from 
above. Combining the analyses of the Subsections 7.2 and 7.3 we conclude that 
the value of $E_{max}$ for, e.g.,  the spectrum $N\propto \sqrt{E}$ should lie 
between $35$ MeV and  $150$  MeV in order not to exceed either the line or 
continuum fluxes. 
 
\begin{figure}[thbp]  
\mbox{
\includegraphics[width=1.\hsize,clip]{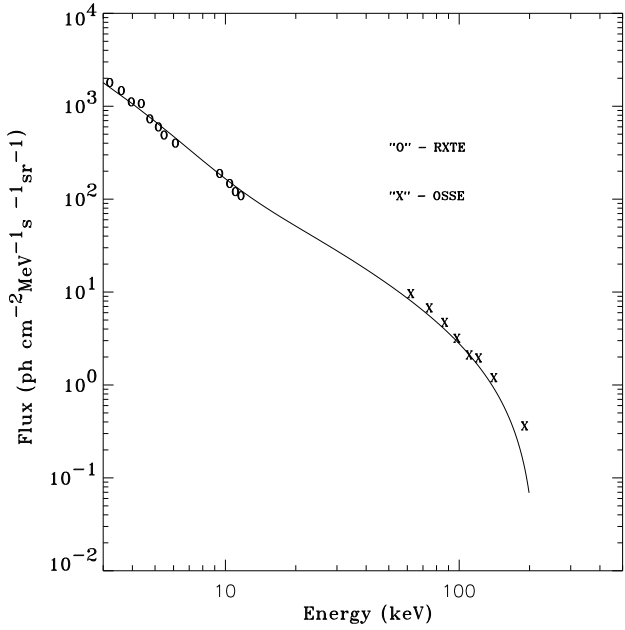}
}     
\caption {The X-ray emission produced by a proton spectrum with a sharp cut-off (step-function) at $400$ MeV.}  
\label{extr}  
\end{figure}

In principle the whole flux of hard X-ray emission up to $200$ keV can be 
produced by proton bremsstrahlung if the cut-off is much steeper than 
exponential  and if the cut-off energy does not exceed the threshold energy for$\pi^0$ production ($\sim 400$ MeV). For an extreme spectrum of protons, 
$N(E)\propto E^{-\gamma^\prime}\theta (400~\mbox{MeV}-E)$ the $\pi^o$ flux  is 
zero but these protons produce X-ray emission up to  $200$ keV (see 
Fig.\ref{extr}). In this case we do not have  problems with the line emission 
since it is only $2.2\cdot 10^{-5} $ ph cm$^{-2}$ s$^{-1}$rad$^{-1}$ for the 
\oxygen line and $8\cdot 10^{-6}$ ph cm$^{-2}$ s$^{-1}$rad$^{-1}$ for the 
\carbon line.           
 
\section{Electron Bremsstrahlung Origin of the Ridge X-Rays} 
The estimates of the parameters for the emitting electrons on the assumption 
that they fill the whole Galactic disk (filling factor $\xi=1$) are presented 
in Table 1.  If we compare these values with those of the protons from Table 1 we find that 
 the energy density of the emitting electrons is much lower than that of the 
 protons. On the other hand, the electrons need more effective acceleration 
 than protons because of the high rate of ionization loss.   
 
The  frequency of electron collisions $\nu_0^{e}$   at  $E=kT$ is: 
\begin{equation} 
\nu_0^e\simeq 2.3\cdot 10^{-10}~\mbox{sec$^{-1}$}\,. 
\end{equation}  
Then from Eq.(\ref{est}) we obtain that $\alpha_0^\prime/\nu_0^e\simeq 
 1.15\cdot 10^{-3}$ or  
\begin{equation} 
\alpha_0^\prime \sim 2.6 \cdot 10^{-13}~\mbox{sec$^{-1}$}~\mbox{for electrons.} 
\label{aed} 
\end{equation}

It is interesting to notice that the time of {\it in 
situ} acceleration  of electrons $(\alpha^\prime)^{-1}$ derived from the  
value 
of the ridge X-ray flux is close to obtained by Schlickeiser (1997) from  
the 
analysis of the hard X-ray flux from the ridge and  electron interactions  
with 
Alfven and Whistler waves in the interstellar medium ($\sim 10^{13}$  
sec). 
 
\begin{figure}[thbp]  
\mbox{
\includegraphics[width=1.\hsize,clip]{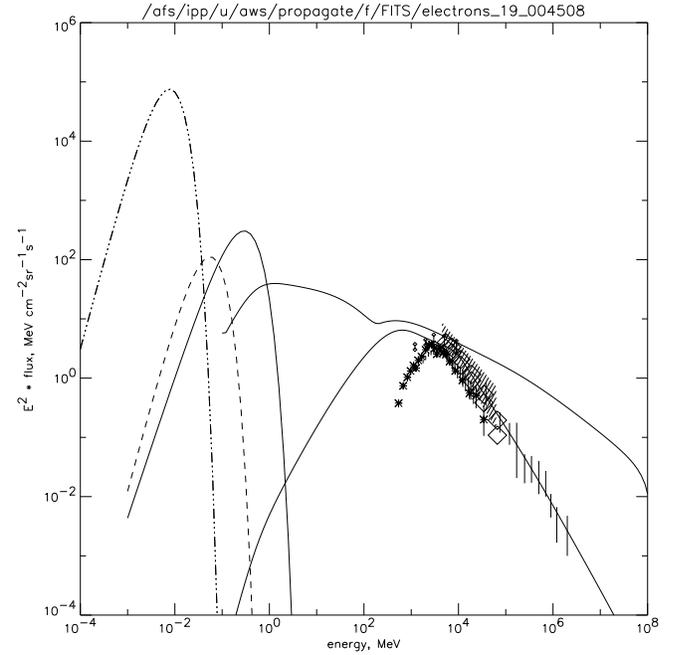}
}   
\caption {The function $E^2\cdot I(E)$ for the observed relativistic and 
derived subrelativistic spectra of  interstellar electrons (solid line - 
Spectrum I, dashed line -  Spectrum II), dashed-dotted-dotted line - the 
thermal spectrum of electrons. In the relativistic energy range measurement of 
the electron flux  at Earth as well as the estimates of interstellar spectra 
derived from the radio and gamma-ray data are shown (for details see Strong et 
al. 2000)).}  
\label{pint}  
\end{figure}

Spectrum I and Spectrum II for the emitting electrons as well as 
 the spectra of relativistic electrons observed near Earth and   the  
thermal 
 spectrum (dashed-dotted-dotted line) of the hot plasma in the  
acceleration 
 regions are shown in Fig.\ref{pint}. If the electron bremsstrahlung 
 interpretation of the ridge emission is correct this figure represents  
the 
 total spectrum of electrons in the Galactic disk from thermal energies    
up to 
the maximum energy of electrons observed near Earth.

The question  arises in the case of an electron bremsstrahlung origin 
of the ridge emission: why is the proton acceleration  suppressed so 
strongly, and under what special conditions  can interactions of background    
electrons with electromagnetic fluctuations  be more effective than those  
of  
protons. Analysis of concrete acceleration processes is beyond the scope of this paper.  
Nevertheless, below we present  
speculations which may explain this effect. 
 
The proton-electron ratio  in the flux of accelerated particles was mainly analysed 
for the case of shock acceleration. It was shown (see e.g. Berezinskii et al. 1990) 
that for the same injection power the flux of protons in the relativistic    
energy range is much higher than that of electrons. This nicely explains the observed 
electron-proton ratio in the flux of the galactic CRs generated by SN shocks. 
  
The situation may be different at subrelativistic energies where the proton Larmor radius is much larger than  
that of electrons.  We notice that the particle Larmor radius  is an essential parameter  
which  affects CR acceleration and propagation (see, e.g., Berezinskii et al. 1990).  
Let us consider a simple situation of particle interaction with magnetic clouds.   
It is clear that this interaction is effective if the particle Larmor radius is smaller than the size of the clouds. Otherwise the  efficiency of particle scattering by the clouds decreases with the particle energy and as a result these  
particles escape almost  
freely from  the acceleration region. As an example we can refer to the paper of Dogiel et  
al. (1987)  where the theory of particle acceleration by neutral gas turbulence was developed. This situation may occur  in  
 the Galactic disk where the degree of ionization 
 is not high and turbulent motions are observed (see Larson 1979). 
 
Dogiel et al. (1987) showed that  stochastic acceleration effectively generates energetic 
particles whose Larmor radius is less than the characteristic correlation length 
of the magnetic turbulence determined by dissipative processes. If the particle energy reaches a value  
where  the Larmor radius is of the order of the  
correlation length  any acceleration is stopped since   
the scattering by the magnetic fluctuations is unable to keep the particles in   
the acceleration region.  
 
\begin{figure}[thbp]  
\mbox{
\includegraphics[width=1.\hsize,clip]{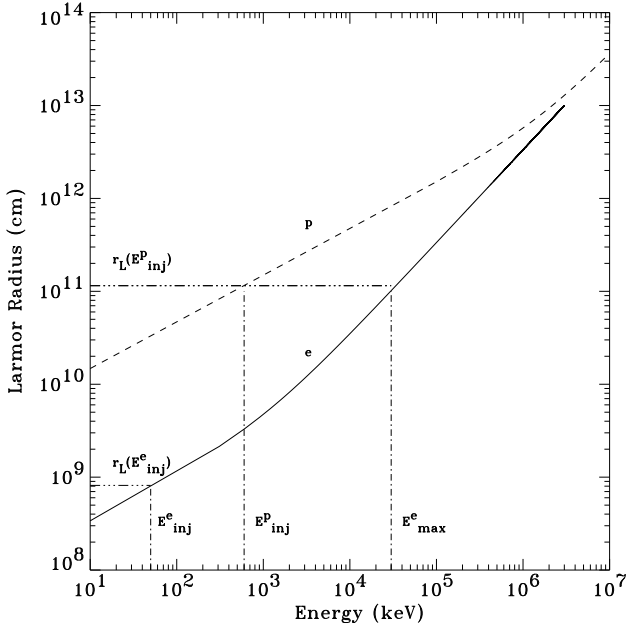}
}   
\caption {Larmor radii of electrons (e) and protons (p) as a function of their  
energy. Vertical lines mark injection enrgies of electrons (at 50 keV) and protons  
(at 600 keV)} 
\label{larm} 
\end{figure}  
 
The energy range of accelerated particles is determined by the inequality $E>E_{inj}$ 
where the injection energy can be estimated from Eq.(\ref{einj}). 
The injection energy estimated from the acceleration frequency (\ref{aed}) equals $E^e_{inj}\sim 50$ keV for electrons and  
$E^p_{inj}\sim 600$ keV for protons.   In Fig.\ref{larm} 
we present Larmor radii $r_L$ of electrons (e) and protons (p) in the interstellar magnetic field $H\sim 3\cdot 10^{-6}$ G    
as a function of their energy. It is clear that if the  
correlation length $L_{cor}$  of magnetic fluctuations  
excited  
by the neutral gas turbulence lies in the interval: $r_L(E^e_{inj})\ll L_{cor}\ll r_L(E^p_{inj})$ then only  
electrons are accelerated while the proton distribution  is an equilibrium Maxwellian. As we see from Fig. \ref{larm} the maximum  
energy of  
electrons in the situation of predominant electron acceleration can be as large as several tens of MeV. 
 
These illustrative arguments present only a speculative possibility  for  predominant electron acceleration and careful analysis  
of the situation will be necessary.  
 
 

\section{Ionization of the Interstellar Medium by Subrelativistic Particles} 
 
Another possibility to find `traces' of particle acceleration in the 
 interstellar medium is an analysis of the ionization state of the background 
 gas since subrelativistic CRs effectively ionize it.  This problem was 
 discussed in Skibo et al. (1996) and Valinia and Marshall (1998). Below we 
 estimate the rate of ionization for the spectra of protons and  electrons 
 derived from the X-ray data. The cross-section of the ionization process has 
 the form (Spitzer and Tomasko 1968) 
\begin{eqnarray}  
\label{si} 
\sigma_i&&\simeq 1.23\cdot 10^{-20}{Z^2\over \beta^2}\cdot\\ 
&&\left( 6.2 +  
\lg\left({\beta^2\over{1-\beta^2}}\right)-0.43\cdot\beta^2\right)\mbox{(cm$^2$)}\,. 
\nonumber 
\end{eqnarray} 
Then the ionization rate $\zeta$ of the medium can be determined from 
\begin{equation} 
\zeta\simeq {\int\limits_E}\sigma_i v {{dN}\over{dE}}dE~\mbox{(sec$^{-1}$)}\,. 
\end{equation} 
We see that the ionization rate depends on the particle velocity $v$ and CR 
 density $N$. As  was mentioned before the values are the same for  X-ray 
 emitting protons and electrons and therefore $\zeta_i^e\simeq \zeta_i^p$. The 
 value of $\zeta$ derived from the observed ridge X-ray flux for different 
 values of the particle spectral index $\gamma^\prime$ is in the range 
 $(8-30)\cdot 10^{-15}$ sec$^{-1}.$ 
 
 Observations of the diffuse Balmer lines give 
  estimates for $\zeta$ in the interstellar medium  which, depending on the 
 observation direction, is between $3\cdot 10^{-15}$ sec$^{-1}$ and $30\cdot 
 10^{-15}$ sec$^{-1}$ (Reynolds et al. 1973); this is close to our estimates.  
 
The source of ionization (as well as the interstellar gas heating) is an old 
 problem (for  reviews on this subject see Dalgarno and McCray 1972, and 
 Spitzer and Jenkins 1975). The energy required to heat the interstellar medium 
 is about $(3-7) \cdot 10^{41}$ erg sec$^{-1}$ and the required energy to keep 
 the medium ionized is of the order of $(2-15)\cdot 10^{41}$ erg sec$^{-1}$. 
 
A central problem of these processes is the unobserved energetic radiation 
 which maintains the heating and ionization state of the interstellar gas.    A 
potential source which in principle could deposit  significant power into the 
interstellar gas was considered  to be either soft X-rays (Silk and Werner 
1969) or a flux of low MeV cosmic rays (Hayakawa et al. 1961, Spitzer 1968). 
The hypothetical flux of these cosmic rays was supposed to be in the form of 
protons with energies 2 -- 5 MeV   (see  Dalgarno and McCray 1972,  Spitzer and 
Jenkins 1975, Nath and Biermann 1994) or in the form of subrelativistic electrons (see Sacher and 
Sch\"onfelder 1984). However, at that time there were no observational data 
which allowed to prove any of these hypotheses.   At the present time the 
situation is more promising. 
 
 The recent observations of the $H_\alpha$-line emission from Milky Way and 
other galaxies are inconsistent with  the pure photoionization model (Reynolds 
et al. 1999). In addition to photoionization   another source is required, 
e.g.,     a flux of subrelativistic particles (Reynolds et al. 1999).  
 
The observed flux of nonthermal X-rays coming from the Galactic disk 
suggests that there are such subrelativistic particles, although it is still 
unclear  whether the flux consists of electrons or protons and what is the 
source of these particles.

\section{Conclusions}

 From the observational data obtained with the GINGA, RXTE and ASCA telescopes 
it definitely follows that the flux of hard X-ray emission is  nonthermal, 
diffuse and is  probably generated by fast particles accelerated from the thermal pool. 
Analyses of processes which can be responsible for this emission suggest only 
two of them, bremsstrahlung radiation of nonthermal electrons or protons. From 
the kinetic equations describing particle acceleration from the thermal pool we 
have calculated the spectra of electrons and protons which can produce the 
observed X-ray flux, estimated the parameters of the acceleration process, and 
determined the chemical composition of the accelerated flux, which is poor in 
heavy elements. We have shown that the spectrum of the emitting particles 
should be  hard in order  to reproduce the observational data. These type of 
spectra are expected from the analysis of processes of {\it in situ} 
acceleration and the subsequent transformation of these spectra in the 
interstellar medium. 
 
In the case of a proton bremsstrahlung origin of the ridge emission the maximum 
energy of the protons is of the order of $150-500$ MeV if the whole range of 
nonthermal X-ray emission up to $200$ keV is produced by the accelerated 
protons. An advantage of this model is the relatively `soft' acceleration 
conditions needed to produce the necessary proton flux. The acceleration time 
in this case is about $2\cdot 10^{14}$ sec. The nuclear component with charge $Z>1$ is 
strongly suppressed.  Our analysis has shown that the proton bremsstrahlung model is free from the previously  discussed problems of the high gamma-ray  flux at MeV or at hundreds of MeV energy regions. However, the model with proton bremsstrahlung requires specific restrictions  on the proton spectrum. Otherwise, the model fluxes of nuclear gamma-ray line or $\pi^o$ photons  exceed  the observational limits. 
The line flux produced by the accelerated protons is below the OSSE and COMPTEL upper limit if the spectrum is very hard and has a cut-off at  energies of  hundreds of MeV. To avoid another problem, the high flux of $\pi^o$ gamma-ray photons, we have to assume that the cut-off is extremely steep and its energy is  just near the threshold energy for $\pi^o$ photon production.

The real problem of the proton bremsstrahlung model is the pressure of the accelerated protons in the disk, which  as follows from Table 1 is 
 very high, but whether or not it gives a rise to hydrodynamical motions in the 
 Galaxy  depends on the degree of coupling between the interstellar gas and the 
 subrelativistic cosmic rays.

For the electron origin of the ridge flux we do not have the problems of the 
 gamma-ray line and $\pi^o$ emission. On the other hand the electron bremsstrahlung model requires more effective acceleration of background particles with  a characteristic time $\sim 10^{13}$ s.  
The  acceleration of protons must 
 be strongly suppressed, although we know from observational data and 
 theoretical analyses that usually the flux of accelerated protons is  larger 
 than that of accelerated electrons,  at least in the relativistic energy range\footnote{We notice, however, that for the case of shock acceleration the density of subrelativistic electrons is higher than the density of protons (Baring et al. 1999)} . We have pesented here arguments in favour of  
predominant electron acceleration in the subrelativistic energy range  
though they cannot be considered as absolutely conclusive. 
 
In the case of an electron bremsstrahlung origin of the ridge X-ray flux a complete spectrum of electrons in the interstellar medium can be derived from the radio, X-ray and gamma-ray data. It  extends from thermal keV energies up to $\sim 10^3$ GeV.

The problem of the electron bremsstrahlung model is that these electrons should produce  narrow 6.4 keV iron line emission  from the disk whose average flux according to the derived electron spectrum  we estimate at $0.1$ ph cm$^{-2}$s$^{-1}$sr$^{-1}$. This flux can be distinguished by the ASCA telescope and would have to be seen in the ridge spectrum as an narrow and prominent singuliarity (see Valinia et al. 2000b). However, in the Galactic ridge the 6.4 keV line was observed only from a rather small region near the galactic center in the directions of two complexes of molecular gas and its width significantly exceeds  that expected for  electron bremsstrahlung  (Koyama et al. 1996). In other directions this 6.4 keV line has not been  observed at all. Instead, the  6.6 keV iron emission is seen in the spectrum of the ridge. The width of the line in these directions is uncertain but if it is of the order of that  observed for the line in the direction of the Galactic center ($\sim 100$ eV) it 
gives a strong  evidence in favour of a flux of suprathermal ions in the ridge medium  (Tanaka et al. 2000). Whether or not this leads us again to the bremsstrahlung proton model, it is a goal of future analyses.

The enigma of the Galactic ridge emission is that we do not see evident sites 
 of the particle acceleration which could supply the necessary energy flux of 
 the order of $10^{42}$ erg s$^{-1}$ needed for the production of the X-ray 
 flux. If the other models of the Galactic ridge X-ray emission can be simply 
 rejected by the observations, the bremsstrahlung model survives but with huge 
 energetic problems. We can speculate only that there may be regions in the 
 Galactic disk where the energy of hundreds or thousands of SN   (as in 
 superbubbles) or OB stars is or was accumulated and relatively recent energy 
 fluctuations in these regions   lead to this X-ray excess (in this respect see, 
 e.g. Kn\" odlseder 2000). Then the observed hard X-ray flux may be interpreted 
 as an afterglow of energetic events which took place in the Galactic disk in 
 the past and which generated a high level of turbulence there. In this case 
 the subrelativistic protons  seem  preferable to electrons since they have a 
  longer lifetime. In favour of the proton bremsstrahlung 
 hypothesis we could mention the marginal detection of the C  
and   
 O gamma-ray line emission from the Galactic ridge  with the COMPTEL telescope, 
 since in the case of the proton bremsstrahlung origin the X-ray flux should be 
 accompanied by such gamma-ray line emission. However, this marginal detection 
 needs confirmation.        
 
On the observational front it should be mentioned that ESA's gamma-ray 
 observatory INTEGRAL is due to be launched in 2002 and will provide precisely 
the large-scale mapping and spectral information in the energy range 20 keV - 
few MeV required to help resolve these issues. 
 
In conclusion we confirm   the importance of the analysis of the X-ray diffuse 
emission in the hard energy range. This analysis covers the gap between the 
non-thermal gamma-ray emission and the thermal soft X-ray emission and thus 
gives  information about the connections between processes in the background plasma and the  
acceleration of nonthermal particles. We cannot state that our analysis has 
produced a definite answer on the origin of the hard X-ray emission. 
Nevertheless, we have formulated the general conditions under which this 
emission can be generated  and estimated parameters of the processes which may 
produce the necessary fluxes of subrelativistic particles that in our opinion 
is  a necessary step towards the solution of the problem.

\begin{acknowledgements}  
 
The authors would like to thank Drs. M.G.Baring, H.Kaneda, A.Valinia and N.Y.Yamasaki for their 
comments, and Dr.R.L.Kinzer who kindly sent to us 
the ps-file of Fig.1. 
 
VAD acknowledges financial support  from the Alexander von Humboldt-Stiftung 
which was very essential for these collaborative researches.  
 
This work was prepared during his visit to Max-Planck-Institut f\"ur  
extraterrestrische Physik (Garching) and he is grateful to his colleagues from 
  this institute for helpful and fruitful discussions.  
 
The final version of the paper was partly done at the Institute of Space and  
Astronautical Science (Sagamihara, Japan). VAD thanks his colleagues from the institute and especially Prof.H.Inoue for their warm hospitality and discussions. 

\end{acknowledgements}  
  
{}

\end{document}